# High-performance non-Fermi-liquid metallic thermoelectric materials

Zirui Dong[1,7], Yubo Zhang[2,3,7], Jun Luo[1,4*], Ying Jiang[4], Zhiyang Yu[5], Nan Zhao[3], Liusuo Wu[3], Yurong Ruan[2], Fang Zhang[2], Kai Guo[6], Jiye Zhang[1], Wenqing Zhang[2,3*]

[1] School of Materials Science and Engineering, Shanghai University, Shanghai 200444, China

[2] Department of Materials Science and Engineering, Department of Physics, and Shenzhen Institute for Quantum Science and Engineering, Southern University of Science and Technology, Shenzhen 518055, China

[3] Shenzhen Municipal Key-Lab for Advanced Quantum Materials and Devices, and Guangdong Provincial Key Lab for Computational Science and Materials Design, Southern University of Science and Technology, Shenzhen 518055, China

[4] Materials Genome Institute, Shanghai University, Shanghai 200444, China

[5] State Key Laboratory of Photocatalysis on Energy and Environment, College of Chemistry, Fuzhou University, Fuzhou 350002, China

[6] School of Physics and Materials Science, Guangzhou University, Guangzhou 510006, China

[7] These authors contributed equally: Zirui Dong, Yubo Zhang.

* Correspondence to: junluo@shu.edu.cn (J.L.); zhangwq@sustech.edu.cn (W.Z.)

**Abstract**

**Searching for high-performance thermoelectric (TE) materials in the paradigm of narrow-bandgap semiconductors has lasted for nearly 70 years and is obviously hampered by a bottleneck of research now[1]. Here we report on the discovery of a few metallic compounds, $TiFe_xCu_{2x-1}Sb$ ($x$ = 0.70, 0,75, 0.80) and $TiFe_{1.33}Sb$, showing the thermopower exceeding many TE semiconductors and the dimensionless figure of merits $zT$s comparable with the state-of-the-art TE materials. A quasi-linear temperature ($T$) dependence of electrical resistivity in 2 K - 700 K and the logarithmic $T$-dependent electronic specific heat at low temperature are also observed to coexist with the high thermopower, highlighting**

**the strong intercoupling of the non-Fermi-liquid (NFL) quantum critical behavior[2] of electrons with TE transports. Electronic structure analysis reveals the existence of fluctuating Fe-$e_g$-related local magnetic moments, Fe-Fe antiferromagnetic (AFM) interaction at the nearest 4c-4d sites, and two-fold degenerate $e_g$ orbitals antiferromagnetically coupled with the dual-type itinerant electrons close to the Fermi level, all of which infer to a competition between the AFM ordering and Kondo-like spin compensation as well as a parallel two-channel Kondo effect. These effects are both strongly meditated by the structural disorder due to the random filling of Fe/Cu at the equivalent 4c/4d sites of the Heusler crystal lattice. The temperature dependence of magnetic susceptibility deviates from ideal antiferromagnetism but can be fitted well by $x(T) = 1/(\theta + BT^\alpha)$, seemingly being consistent with the quantum critical scenario of strong local correlation as discussed before[3-5]. Our work not only breaks the dilemma that the promising TE materials should be heavily-doped semiconductors, but also demonstrates the correlation among high TE performance, NFL quantum criticality, and magnetic fluctuation, which opens up new directions for future research.**

TE materials have been researched and developed for 200 years since the Seebeck effect was discovered in 1821. Due to the promising applications of TE materials in waste heat power generation and solid-state refrigeration, persistent efforts have been made to improve TE performance. Metals were studied as TE materials firstly, but they are no longer considered as good TE materials because of their small Seebeck coefficients and low $zT$ values ($zT = S^2\sigma T/\kappa$, where $S$, $\sigma$, $T$, and $\kappa$ are the Seebeck coefficient or thermopower, electrical conductivity, absolute temperature, and total thermal conductivity, respectively). Instead, the best TE materials at present, such as $Bi_2Te_3$[6], $PbTe$[7], $GeTe$[8], and $CoSb_3$[9], are all heavily-doped narrow-bandgap semiconductors. Thus, the currently common consensus in the TE field is that the first-rank TE material should be a heavily-doped narrow-bandgap semiconductor[10], which can achieve the optimal power factor $S^2\sigma$ by balancing the Seebeck coefficient and electrical conductivity[11]. As a result, an optimized $zT$, which determines the TE conversion efficiency, is realized.

After about 70 years of development, the only commercialized TE material is the $Bi_2Te_3$-based material with a peak $zT$ around the unity. In addition, reliably reproducible $zT$ is around 2.0, the reported $zT$ values above 2.0 are often in debate, and all belong to heavily-doped narrow-bandgap semiconductors[1,12,13]. It seems desperately needed to break through the limitation of narrow-bandgap semiconductors for discovering novel TE materials. Many attempts have been made to explore non-semiconductor TE materials. High spin entropy was once believed to enhance the TE performance of a few lamellar cobalt oxides[14-16]. For instance, $Na_xCo_2O_4$ with $Co^{3+}/Co^{4+}$-determined spin entropy achieves Seebeck coefficients about 100 and 200 μV K$^{-1}$ at 300 K and 800 K, repectively[17,18]. Up to now, the maximum $zT$ value of this type of oxide approaches ~1.0 at 800 K after a long-time endeavor[19], halted in thermodynamics by the spin entropy limit from $d$-band degeneracy.

Over the years, a few rare-earth 4$f$-electron-based heavy-fermion systems were found to show metallic electrical conductivities and relatively large Seebeck coefficients[20] at extremely low temperatures. There also exist a few exceptions to the recognized heavy-fermions as $YbAl_3$[21] and $CePd_3$[22], showing thermopower as high as ~100 μV K$^{-1}$ at room temperature. Arguably, this was also considered as hybridizing $f$ electrons with conduction band, resulting in a resonant peak near the Fermi surface[23,24] as in normal materials. The peak $zT$ values of $CePt_3$[22] and $YbAl_3$[21] reach about 0.2 and 0.3 at 300 K, respectively. Nevertheless, the optimal Seebeck coefficients for both systems reach only < 110 μV K$^{-1}$ and impede further enhancement of their TE performance. Very recently, the spin fluctuation was reported to enhance the thermopower of weak itinerant ferromagnetic alloys $Fe_2V_{0.9}Cr_{0.1}Al_{0.9}Si_{0.1}$ and $Fe_{2.2}V_{0.8}Al_{0.6}Si_{0.4}$[25] around the Curie temperature $T_C$. A broad shoulder-like hump on the $S(T) \sim T$ curve was observed around $T_C$, leading to only a 15% to 20% enhancement of the Seebeck coefficient at $T_C$. The well-known NFL superconducting oxides, such as $YBa_2Cu_4O_8$[26] and $La_{2-x}SrCuO_4$[27], also show certain intriguing TE behavior, but these oxides always have negligible $zT$ values, several orders of magnitude lower than the state-of-the-art TE semiconductors.

Conceptually, there is no substantial breakthrough in the search for non-semiconductor TE materials. In this work, we find that a few Heusler-like materials, TiFe$_x$Cu$_{2x-1}$Sb ($x$ = 0.70, 0,75, 0.80) and TiFe$_{1.33}$Sb with excess Fe/Cu occupying the vacant sites of the half-Heusler lattice, show peculiar NFL metallic transport together with excellent TE performance. A high power factor of 20.8 μWcm$^{-1}$K$^{-2}$ and a $zT$ value of 0.75 are achieved in TiFe$_{0.7}$Cu$_{0.4}$Sb at 973 K, and the other systems also show $zT$s higher than 0.3, all comparable with the best half-Heusler TE semiconductors.

**Unconventional metallic transport properties**

Figure 1 plots the resistivity $\rho(T)$ of TiFe$_x$Cu$_{2x-1}$Sb and TiFe$_{1.33}$Sb. They all show quasi-linear dependence in the range from near-zero temperature up to 500 K, indicating that both materials are unconventional metals. Another typical characteristic of metallicity is the vanishing small Seebeck coefficients, $S(T)$, when approaching 0 K (the inset of Fig. 1). The Seebeck coefficients of the TiFe$_x$Cu$_{2x-1}$Sb and TiFe$_{1.33}$Sb gradually increase and become comparable to those of heavily-doped narrow-bandgap semiconductors at 300 K and higher. Moreover, the quasi-linear- or linear-$T$ dependence of $\rho(T)$ in the above materials is valid in a wide temperature range (0 – 500 K in Fig. 1; see also Supplementary Fig. 1a for a wider range of 0 – 973 K) while stably keeping the thermopower as high as approaching 200 μV K$^{-1}$. By fitting to the low-temperature resistivity with the Bloch–Grüneisen model[28] (Supplementary Information section 1 and Supplementary Fig. 2 and Table 1), the parameter of Debye-temperature term is estimated to be as low as ~40 K, much lower than the phonon-mediated Debye temperature of 389 K estimated from the measured sound velocity (Supplementary Information section 2 and Supplementary Table 2). X-ray photoemission spectroscopy (XPS) experiments reveal that the Fe element in both TiFe$_{0.7}$Cu$_{0.4}$Sb and TiFe$_{1.33}$Sb has a valence of zero, and their Fe-2$p$ XPS spectra are almost identical to the metallic Fe, FeSe and FeTe but are obviously different from FeO and Fe$_2$O$_3$ (Supplementary Fig. 3).

The resistivity of normal or conventional metals shows the $\rho(T) \sim T^2$ dependence in Landau-Fermi-liquid paradigm, in contrast, the linear and quasi-linear temperature dependence of resistivity [i.e., $\rho(T) \sim T$] is usually considered as a canonical signature of the NFL metals. Notice also that the resistivities of the above materials, $\rho(T) \sim 10^{-6} - 10^{-5}$ Ω m, are one to two orders of magnitude higher than the constantan[29] ($Cu_{0.6}Ni_{0.4}$, an alloy of copper and nickel), the best TE metal. The electrical mean-free paths could be estimated to be as low as 0.47 nm in $TiFe_{0.7}Cu_{0.4}Sb$ and 0.73 nm in $TiFe_{1.33}Sb$ (Supplementary Information section 3 and Supplementary Table 3), which are close to or even smaller than the lattice constants. All the above observations point to the bad metallicity and unconventional metallic transport in $TiFe_{0.7}Cu_{0.4}Sb$ and $TiFe_{1.33}Sb$.

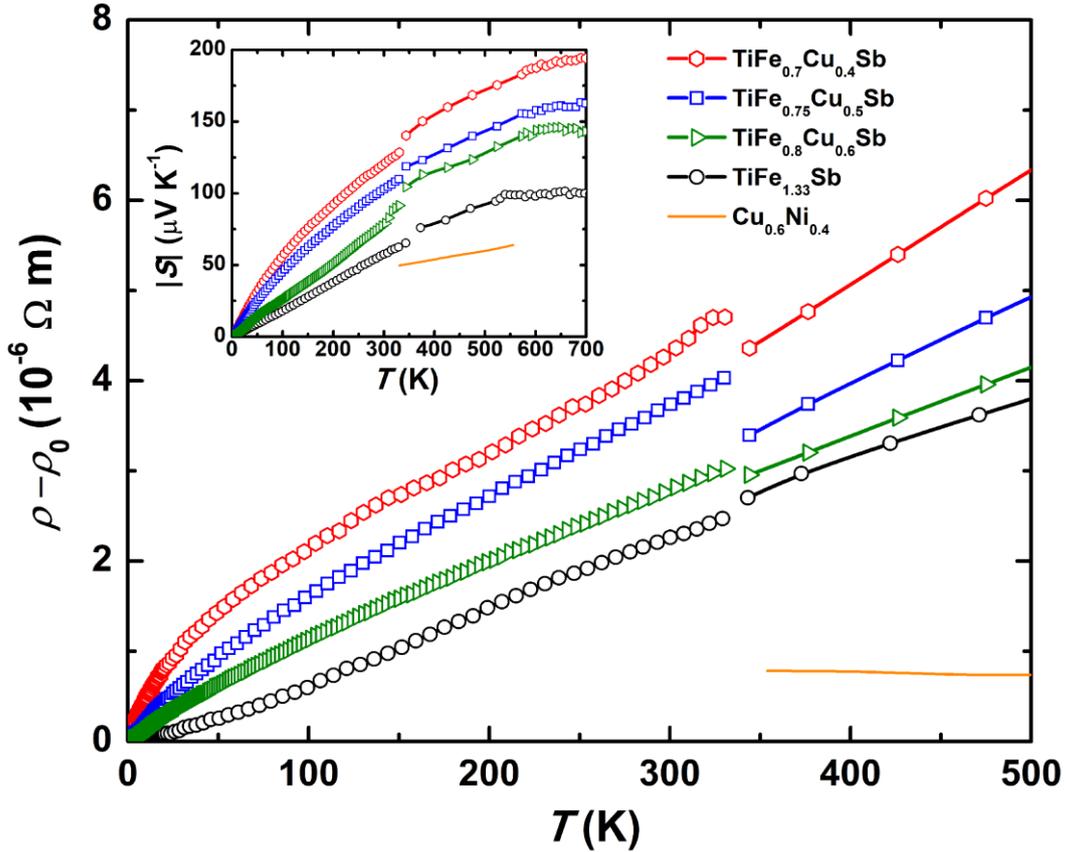

**Fig. 1 Temperature-dependent electrical resistivities and Seebeck coefficients of $TiFe_xCu_{2x-1}Sb$ and $TiFe_{1.33}Sb$ samples.** The data of the $Cu_{0.6}Ni_{0.4}$ alloy (solid orange line) are included for comparison[29]. $\rho_0$ is the resistivity at 0 K, which is obtained by fitting to the low-temperature resistivity using the Bloch–Grüneisen model (Supplementary Information section 1 and Supplementary Fig. 2 and Table 1). The resistivity of $Cu_{0.6}Ni_{0.4}$ shown in this figure is the raw resistivity without eliminating the $\rho_0$ part.

**Exceptional thermopower and TE performance**

Despite the metallicity, the above materials exhibit excellent TE properties (Fig. 2). The thermopower increases continuously with the rising of temperature up to 600 K. The Seebeck coefficient of TiFe$_{0.7}$Cu$_{0.4}$Sb reaches 194 μV K$^{-1}$ at 700 K, the highest among all known TE metals, with the reported peak value of ~60 μV K$^{-1}$ for constantan[29], ~-110 μV K$^{-1}$ for YbAl$_3$[21], and ~110 μV K$^{-1}$ for CePd$_3$[22]. There appears to be a nonlinear turnover above 600 K, which could be ascribed to the structural transition (Supplementary Fig. 4), but all the materials still maintain high values of thermopower (Supplementary Fig. 1b) while simultaneously a reasonably low resistivity (Supplementary Fig. 1a) at a temperature close to 1000 K. Coupled with the very low thermal conductivity (Supplementary Fig. 5), the metallic TiFe$_x$Cu$_{2x-1}$Sb and TiFe$_{1.33}$Sb show remarkable TE performances, and the TiFe$_{0.7}$Cu$_{0.4}$Sb reaches a $zT$ value of 0.75 at 973 K, comparable with or even better than those of typical half-Heusler TE semiconductors[30-34]. This value of $zT$ is recognized to be the highest one among all the ternary Heusler or Heusler-like materials including those half-Heusler TE semiconductors optimized only at the transition metal (Fe/Co/Ni) site[31-34].

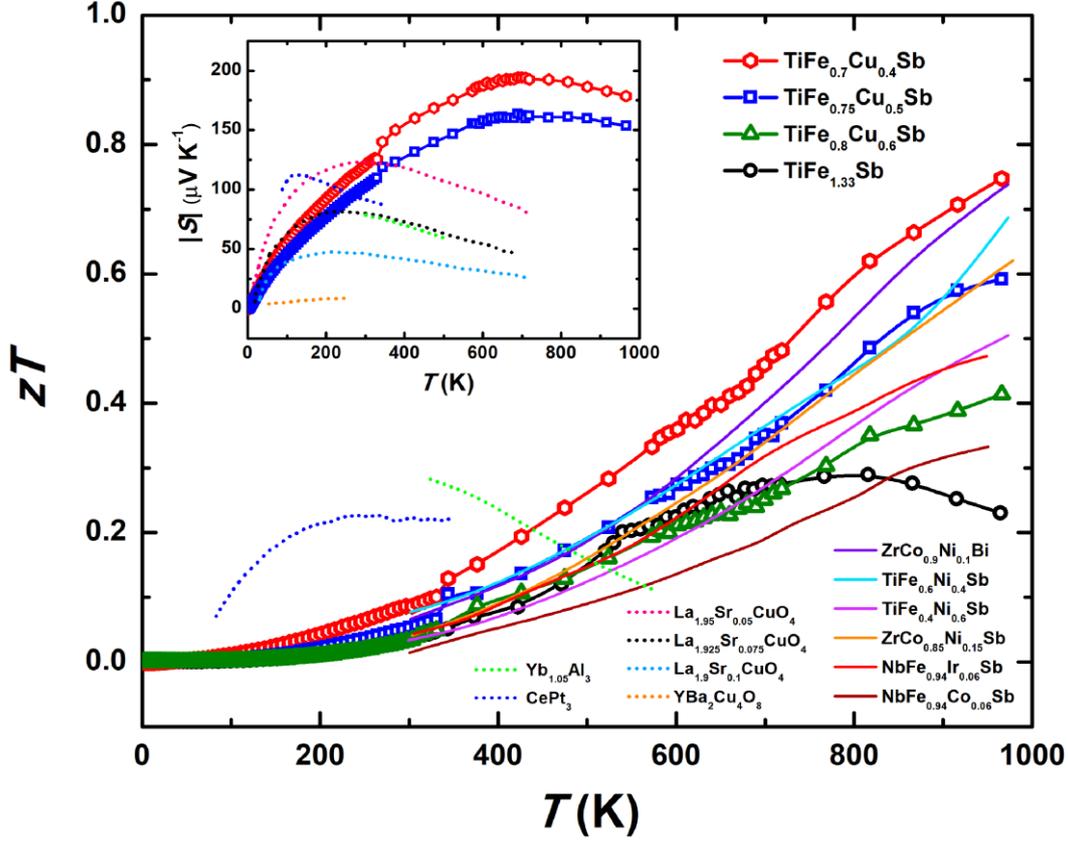

**Fig. 2 Temperature-dependent $zT$ values and Seebeck coefficients of $TiFe_xCu_{2x-1}Sb$ and $TiFe_{1.33}Sb$, all in lines with symbols.** The $zT$ values of $CePt_3$, $Yb_{1.05}Al_3$, $ZrCo_{0.9}Ni_{0.1}Bi$, $TiFe_{0.6}Ni_{0.4}Sb$, $TiFe_{0.4}Ni_{0.6}Sb$, $ZrCo_{0.85}Ni_{0.15}Sb$, $NbFe_{0.94}Ir_{0.06}Sb$, and $NbFe_{0.94}Co_{0.06}Sb$, and the Seebeck coefficients of $CePt_3$, $Yb_{1.05}Al_3$, $La_{1.95}Sr_{0.05}CuO_4$, $La_{1.925}Sr_{0.075}CuO_4$, $La_{1.9}Sr_{0.1}CuO_4$, and $YBa_2Cu_4O_8$ are taken from the literature for comparison[21,22,26,27,30-34], all in lines without symbols.

**Logarithmic temperature-dependent electronic specific heat**

To elucidate the origin of the exceptional TE properties of the metallic $TiFe_xCu_{2x-1}Sb$ and $TiFe_{1.33}Sb$ materials, low-temperature specific heat and resistivity were also measured (Fig. 3). Interestingly, the temperature-dependent electronic specific heat $C_{el}$ of both $TiFe_{0.7}Cu_{0.4}Sb$ and $TiFe_{1.33}Sb$ follow the relationship $C_{el}/T \sim -\ln T$ (see Supplementary Information section 4 and Supplementary Fig. 6 and Table 4 for details of $C_{el}$ extraction), while the resistivities of those systems still show nearly linear dependence on the temperature. The observed coexistence of the quasi-linear temperature dependence of resistivities and the logarithmic temperature-dependent specific heats strongly hints at the typical features of NFL bahavior[35,36] and quantum critical phenomenon as will be discussed later.

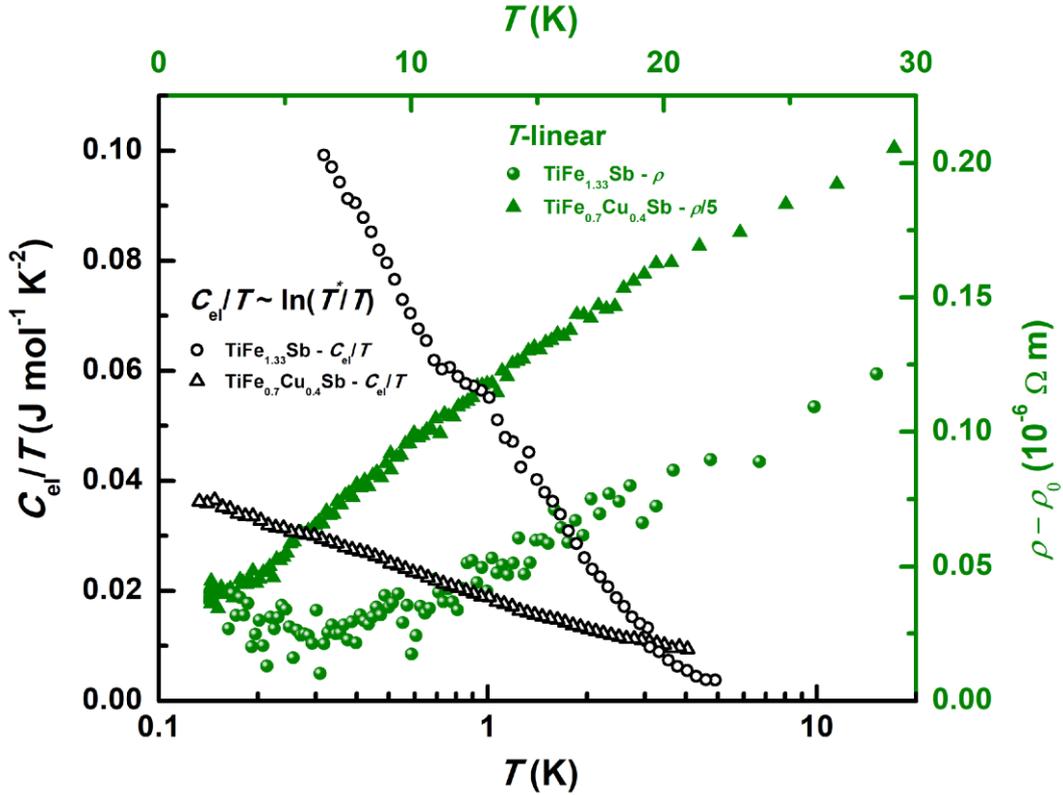

**Fig. 3 Temperature-dependent electronic specific heat capacities and resistivities of the TiFe$_{0.7}$Cu$_{0.4}$Sb and TiFe$_{1.33}$Sb samples.** Note that the electronic specific heat coefficients are at least one order-of-magnitude higher than those of pure metals (for example, Cu[37] and CuNi[38] and even approach those of rare-earth-related heavy Fermi systems[39].)

**Random distribution of Fe/Cu**

All the TiFe$_x$Cu$_{2x-1}$Sb and TiFe$_{1.33}$Sb samples crystallize in a single-phase Heusler-like structure with the space group $F\bar{4}3m$ even with excess Fe/Cu atoms as revealed by the X-ray powder diffraction (XRD) patterns (Supplementary Fig. 7 and 8). The structure of TiFe$_{1.33}$Sb agrees with that of Tavassoli et al[40]. who reported the phase relation and TE properties of Ti$_{1+x}$Fe$_{1.33-x}$Sb. Both the high-angle annular dark-field scanning transmission electron microscopy (HAADF-STEM) image and energy dispersive X-ray spectroscopy (EDS) maps (Fig. 4a) clarify a uniform distribution of compositions in the samples. The aberration-corrected high-resolution HAADF image along the [110] direction reveals that both the 4c and 4d Wyckoff sites in the half-Heusler lattice are partially occupied for TiFe$_{0.7}$Cu$_{0.4}$Sb and TiFe$_{1.33}$Sb (left panels of Fig. 4b and 4c), and the corresponding intensity profiles of HAADF reveal no obvious difference in the atomic occupancies of the 4c or 4d sites (right panels of Fig. 4b and 4c). The integrated differential phase contrast (iDPC) image also indicates that the 4c and 4d sites of TiFe$_{0.7}$Cu$_{0.4}$Sb are nearly randomly occupied (Fig. 4d). All the above observations about structures clearly conclude that the 4c and 4d sites in the Heusler lattice are nearly randomly occupied by the Fe/Cu atoms without noticeable preference. This is reasonable because the 4c and 4d sites in a Heusler lattice could be considered to be equivalent to each other from a pure crystallography point of view, with identical both the nearest neighbors and cubic symmetry.

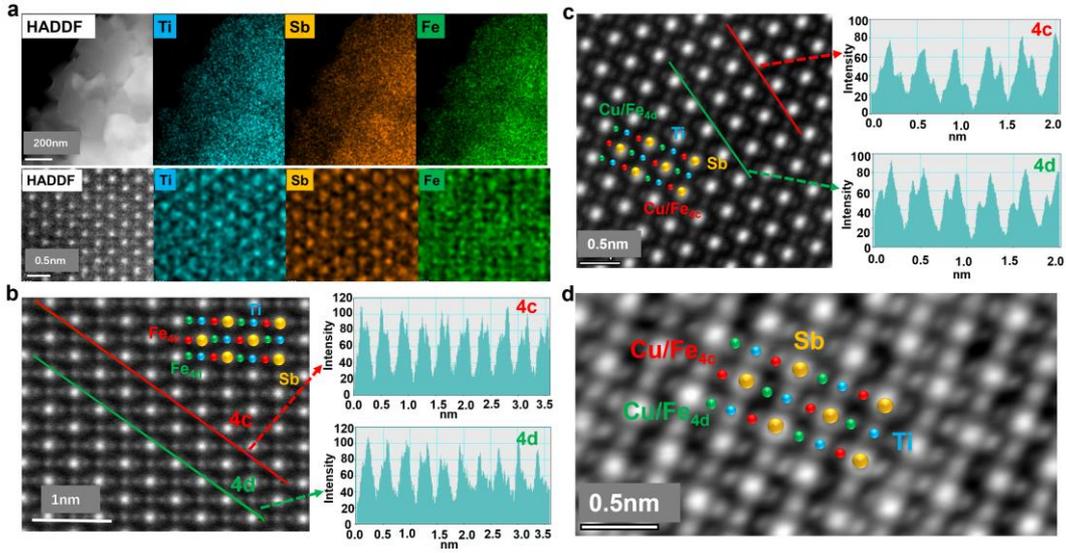

**Fig. 4 Microstructures of TiFe$_{0.7}$Cu$_{0.4}$Sb and TiFe$_{1.33}$Sb. a,** HAADF-STEM image and corresponding EDS maps for TiFe$_{1.33}$Sb. **b, c,** Aberration correction high-resolution HAADF image along a [110] direction and the corresponding intensity profiles across the horizontal lines for TiFe$_{1.33}$Sb and TiFe$_{0.7}$Cu$_{0.4}$Sb, respectively. **d,** iDPC image along a [110] direction of the TiFe$_{0.7}$Cu$_{0.4}$Sb sample.

**Origin of the bad-metallicity and quantum criticality**

The TiFe$_x$Cu$_{2x-1}$Sb and TiFe$_{1.33}$Sb samples show a canonical signature of the NFL metals, i.e., the linear temperature dependence of resistivity [$\rho(T) \sim T$]. While the measured electrical transport implies a breakdown of the quasiparticle-based Fermi-liquid picture, it would be more exciting for the discovery of the NFL TE metals with high $zT$s at (or above) room temperature and its proximity to a quantum critical point with logarithmic-in-$T$ specific heat, which may provide a basis for a possible paradigm shift of research. To get insight into the origin of the exotic physical properties, density-functional-theory (DFT) calculations are carried out to understand the electronic structures of the materials. To include the 4c-4d random occupation, the Fe$^{4c}$-Fe$^{4d}$ disordering is simulated in a superstructure for TiFe$_{1.33}$Sb, in which 216 tetrahedral sites (108 4c and 108 4d) are randomly occupied by 144 Fe ions (Supplementary Fig. S9) based on the above experimental observations. It should also be addressed that different configurations of the superstructure with quasi-random occupancy of the Fe sites lead to almost identical conclusions (Supplementary Fig. 10).

The calculated electronic structures predict TiFe$_{1.33}$Sb as a metal, and the band-edge states around the Fermi level ($E_F$) show a mixture of Fe-3$d$, Ti-3$d$, and noticeable Sb-5$p$ electrons (Fig. 5a). This is understandable because the current materials could be considered as embedding Fe/Cu atoms into the closely packed Ti-Sb matrix. While the interaction in the Ti-Sb matrix is determined by Ti-3$d$ and Sb-5$p$ orbitals, Fe atoms at the 4c-4d sites bond through Fe-3$d$-orbitals with both Ti-3$d$ and Sb-5$p$ orbitals in the matrix. Due to the well-kept cubic symmetry, the Fe-3$d$ states close to the $E_F$ are dominantly the $e_g$ states with 2-fold degeneracy and partially unfilled. Spin-polarized calculations reveal that Fe atoms have a strong magnetic instability (Fig. 5b) to save considerable energies (> 40 meV/Fe atom) compared with the spin-unpolarized case. The Fe-3$d$-based local magnetic moments are relatively diverse, covering from ~zero to values as high as ~1.0 μ$_B$ or even higher (Fig.5b), and fluctuate from site to site. The nominally empty Ti-3$d$ orbitals have small electron occupation and are passively polarized through the hybridization with the Fe-3$d$ orbitals, and show relatively small magnetic moments fluctuating around 0 ~ ±0.4 μ$_B$. It should also be addressed that the band-edge Ti-3$d$ electrons in TE HH semiconductors with Ti-Sb matrix are dominantly itinerant as understood before[33,40]. Sb-5$p$ electrons show itinerant characteristics and are nearly uniformly distributed. Figures 5b and 5c depict the whole picture of the distributions of both fluctuating local magnetic moments and dual-type itinerant electrons around the $E_F$.

There are a few specific aspects that make the current materials intriguing and interesting. Firstly, it is the Fe-$3d$-$e_g$-states with fluctuating local magnetic moments that embed in a metallic environment. The bandwidth of the $e_g$-states close to $E_F$ is found to be substantially narrow, only about ~1.5 eV, due to the much expanded nearest-neighboring Fe-Fe distance (~3 Å in TiFe$_{1.33}$Sb, compared with 2.48 Å in the BCC Fe) with the constraint from the Ti-Sb matrix. The on-site Coulomb interaction U for the $e_g$ electrons is evaluated by the constraint-random-phase-approximation (CRPA) and is around 1.7 eV, clearly implying the role of strong electron correlation. They point to the special role of the localized Fe-$3d$-$e_g$ electrons, and naturally refer to the fact of magnetic impurities in an itinerant electron environment. This has been studied for a long time that the localized states, Fe-$3d$-$e_g$ in the current case, with magnetic moment introduce strong correlation effects and Kondo effect into itinerant electron system and thus the NFL behavior[41,42]. Secondly, there exist symmetry-enforced 2-fold degenerate Fe-$3d$-$e_g$ states coupled with the dual-type itinerant electrons, mixing of the Sb-$5p$ and Ti-$3d$, close to the $E_F$. This is also a strong indication of the multi-channel Kondo effect that has also been investigated early for multiple-orbital magnetic impurities in a metallic environment[43,44]. Both points could lead to quantum criticality. Thus, the magnetic quantum criticality, suggested by the logarithmic-in-$T$ specific heat, also finds microscopic signatures from the simulated electronic properties.

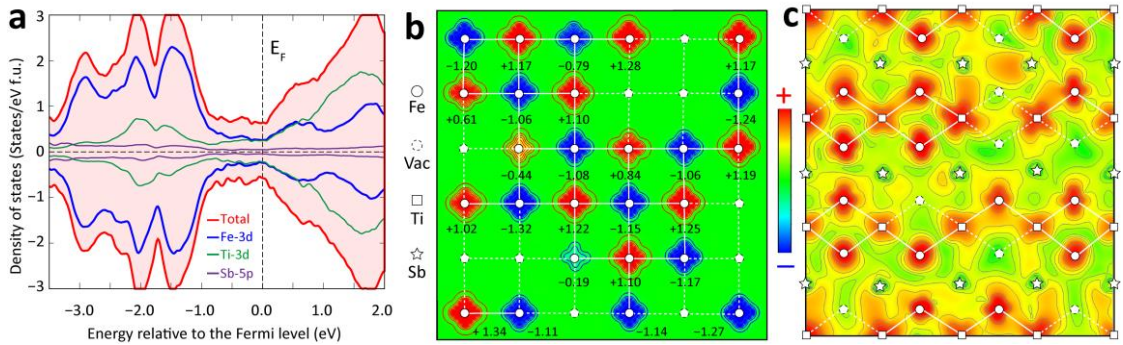

**Fig. 5 Bad metallicity, antiferromagnetism, transport networks of itinerant electrons, and spin fluctuations in TiFe$_{1.33}$Sb from theoretical simulations. a,** Electronic density of states. **b,** Spin density of a (001) plane containing Fe$^{4c}$ and Fe$^{4d}$ atoms with the local magnetic moments denoted by the number below the atoms. A redder (bluer) color denotes a more positive (negative) value, and the rule also applies to subplots **c**. **c,** Electron density on a (110) plane from $E_F$ - 0.1 to $E_F$ + 0.1, where $E_F$ is the Fermi level (spin-up channel only, See Supplementary Fig. 9 for spin-down channel).

These materials show two types of AFM interactions, i.e., $Fe^{4c}$-$Fe^{4d}$ AFM at the nearest-neighbor 4c-4d sites and Fe-Ti AFM, unaffected by the structural complexity due to the 4c/4d occupation disordering. The $Fe^{4c}$-$Fe^{4d}$ AFM (Fig. 5b) tends to form a checkerboard-like magnetic ordering, although being constantly disrupted by the disordered vacancies. Calculations prove that Ti-3$d$ orbitals always develop the opposite spin densities around the nearby Fe-3$d$ ions, providing antiferromagnetically-polarized electrons to form a spontaneous spin compensation. The two AFM interactions also frustrate each other in a $Fe^{4c}$-$Fe^{4d}$-Ti triangular local geometry. The effective Fe-Ti exchange interaction has a relatively strong strength of 30 meV or even higher due to the effective overlap between the itinerant Ti-3$d$ orbitals and the Fe-3$d$ states. Because of the good itinerancy of Ti-3$d$ states, the discussion about the multi-channel Kondo effect takes Ti-3$d$ electrons as heavy itinerant ones, which is considered to be reasonable based on the transport data from many Co-based HH semiconductors[45-49].

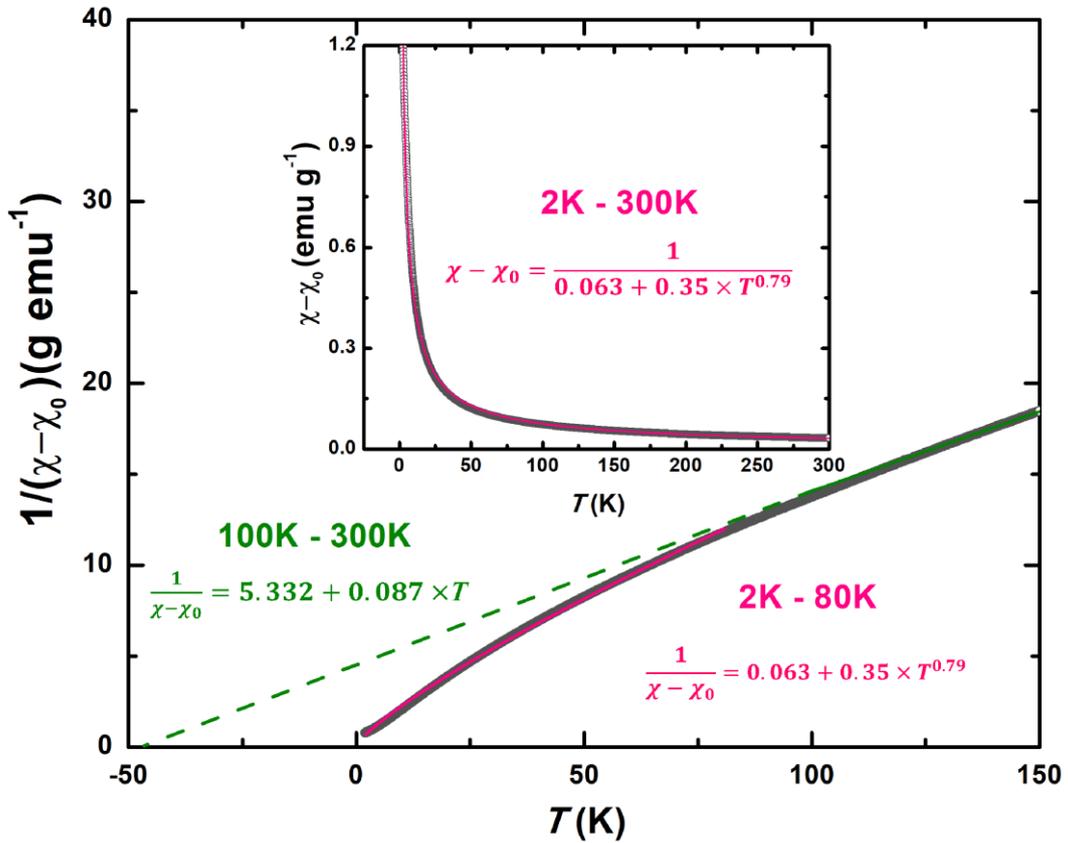

**Fig. 6 Temperature-dependent magnetic susceptibility of TiFe$_{1.33}$Sb.**

Notice that both TiFe$_x$Cu$_{2x-1}$Sb and TiFe$_{1.33}$Sb contains strong disordering of the magnetic atoms, which significantly enhances the fluctuations of the local magnetic moments of Fe atoms. It is thus reasonable to infer that the observed NFL and quantum criticality come from the integrated effects of magnetic-impurity-induced scattering, Fe$^{4c}$-Fe$^{4d}$ AFM ordering at the nearest-neighbor sites, and the multi-channel Kondo effect, and all are strongly mediated by the random distribution of magnetic atoms. It is therefore expected that the NFL and quantum criticality should manifest localization character to some extent, even hardly quantified for now. Low-temperature magnetic susceptibility indicates that the TiFe$_{1.33}$Sb sample is AFM coupling with a Curie-Weiss temperature $T_{CW}$ of −56 K by fitting with the Curie-Weiss law (Supplementary Information section 5 and Supplementary Fig. 11). Nevertheless, the magnetic susceptibility below 100 K offsets from the fitting curve, implying that the normal Curie-Weiss law is not applicable to the sample and unique magnetic and spin fluctuations may exist. Interesting, the magnetic susceptibility of the sample can be well described in the temperature ranges of both 2 K - 100 K and 2 K - 300 K (Fig. 6) by the model $x(T) = 1/(\theta + BT^\alpha)$ established for the system with local quantum criticality[3]. The power index $\alpha$, reflecting the dependence of the magnetic susceptibility on the temperature, is 0.79, agreeing surprisingly well with the value of 0.75 as theoretically derived in the literature[3].

In summary, we discover that the NFL metal showing a quantum critical behavior may be a new type of high-performance TE material. The random filling of Fe/Cu atoms on the equivalent 4c/4d sites of the Heusler metals TiFe$_x$Cu$_{2x-1}$Sb and TiFe$_{1.33}$Sb, leads to the structure disorder, strong magnetic fluctuation, and quantum critical behavior of these non-conventional metals. The Seebeck coefficient of TiFe$_x$Cu$_{2x-1}$Sb and TiFe$_{1.33}$Sb increases continuously with the increase of temperature, while a low resistivity is maintained at the same time. Combined with the very low thermal conductivity due to the disordered structure, the TiFe$_x$Cu$_{2x-1}$Sb and TiFe$_{1.33}$Sb samples achieve the TE performance comparable to or even better than the traditional TE semiconductors. The TiFe$_{0.7}$Cu$_{0.4}$Sb sample shows a Seebeck coefficient of 194 µV/K at 700 K and a *zT* value of 0.75 at 973 K, demonstrating the potential to search for high-performance TE materials along this direction.

**Methods**

**Sample synthesis.** Polycrystalline $TiFe_xCu_{2x-1}Sb$ samples ($x$ = 0.7, 0.75 and 0.8) and $TiFe_{1.33}Sb$ were prepared by high-energy ball milling and spark plasma sintering (SPS). The starting materials (Ti shots 99.6%, Fe pieces 99.99%, Cu shots 99.99%, and Sb shots 99.999%) were weighed according to the nominal composition of the sample and loaded into the Ar-protected stainless-steel jar, which were ball-milled for 30 h using the SPEX 8000M Mixer/Mill. Then, the ball-milled powders were put into a graphite die with an inner diameter of 12.7 mm and consolidated into pellets under 60 MPa by SPS at 973K for 20 min. The relative mass densities of the samples are between 94% and 98% (Supplementary Table 5).

**Structure Characterization.** Phase identification and crystal structure analysis were carried out with high-resolution powder XRD patterns collected by a Rigaku SmartLab-II diffractometer with Cu-K$_\alpha$ radiation. The microstructures of the samples were examined by a high-resolution TEM (JEM-F200, JEOL, Japan) and a probe Cs-corrected TEM (Themis ETEM, Thermo Fisher Scientific, USA). TEM specimens were prepared by mechanical slicing, polishing, and dimpling, followed by ion-milling. Energy-dispersive X-ray spectroscopy was used to determine the distribution of elements at the nanoscale.

**Transport properties measurements.** High-temperature electrical and low-temperature thermal transport properties were performed on a rod-like sample cut directly from the SPS particle. Electrical conductivity ($\sigma$) and Seebeck coefficient ($S$) above room temperature were measured by a four-probe method using a ZEM-3 system (ULVAC-RIKO, Japan). Low-temperature TE properties (5 - 350 K) including $\sigma$, $S$, and $\kappa$ were measured using a physical property measurement system (PPMS, Quantum Design, USA) with the thermal transport option (TTO). The thermal conductivity at high temperature is calculated by $\kappa = \lambda \rho_d C_P$, where $\lambda$ is the thermal diffusivity, $\rho_d$ is the density of the sample, and $C_P$ is the specific heat capacity. $\lambda$ was tested by a laser flash method (LFA 467, NETZSCH, Germany), $C_P$ was estimated by Dulong-Petit law, and $\rho_d$ was measured by the Archimedes method. Room temperature sound velocity measurement by the ultrasonic material characterization System (UMS-100, TECLAB, France). Low-temperature specific heat capacity ($C_P$) was measured by the PPMS with a dilution refrigerator option (PPMS-DR, Quantum Design, USA).

**Density functional theory simulation methods.** The vacancy disordering effect is considered in a 3×3×3 superstructure, in which 216 atomic sites are randomly occupied by 144 Fe ions. Crystal structures are fully relaxed. The DFT calculations are carried out using the SCAN (Strongly Constrained and Appropriately Normed) meta-GGA[50].

The implementation in the VASP[51,52] and FHI-aims[53] codes are crossly checked, and both codes give very similar results. For the VASP simulations, the energy cutoff is 500 eV, and a mesh of 2×2×2 is used for the Brillouin zone integration.

**Data availability**

The data supporting the findings of this study are available from the corresponding author (Jun Luo) upon reasonable request. Source data are provided with this paper.

**Acknowledgments**

This work was supported by the National Key Research and Development Program of China (Nos. 2018YFA0702100 and 2019YFA0704901) and National Natural Science Foundation of China (Grant Nos. 92163212, 51632005, U21A2054, 52072234, and 51772186), and W.Z. also acknowledges the support from the Guangdong Innovation Research Team Project (No. 2017ZT07C062), Guangdong Provincial Key-Lab program (No. 2019B030301001), Shenzhen Municipal Key-Lab program (ZDSYS20190902092905285), and the Centers for Mechanical Engineering Research and Education at MIT and Southern University of Science and Technology, China. Computing resources were supported by the Center for Computational Science and Engineering at the Southern University of Science and Technology. We thank Dr. D.C. Wu at Thermo Fisher Scientific Company for assistance in performing atom-resolved EDS maps.

**Author contributions**

J.L. and W.-Q.Z. conceived and designed the study. Z.-R.D. prepared the samples. Y.-B.Z., Y.-R.R. and F.Z. performed theoretical calculations and physical analysis. Y.J. and Z.-Y.Y. performed the microstructure analysis on TEM. K.G analyzed the crystal structure. Z.-R.D., J.-Y.Z. and J.L. measured and analyzed the electrical and thermal transport properties. Z.-R.D., W.-Q.Z. and J.L. measured and analyzed the low-temperature specific heat capacities. Z.-R.D., N.Z. and L.-S.W. measured and analyzed the magnetic properties. Z.-R.D., Y.-B.Z., J.L. and W.-Q.Z. analyzed the experimental results systemically and co-wrote the paper.

**Competing interests**

The authors declare no competing interests.